# Beta Decay Of Exotic $T_Z = -1$ And $T_Z = -2$ Nuclei

S.E.A.Orrigo[a], B.Rubio[a], Y.Fujita[b], B.Blank[c], W.Gelletly[d], J.Agramunt[a], A.Algora[a], P.Ascher[c], B.Bilgier[e], L.Cáceres[f], R.B.Cakirli[e], H.Fujita[g], E.Ganioğlu[e], M.Gerbaux[c], J.Giovinazzo[c], S.Grévy[c], O.Kamalou[f], H.C.Kozer[e], L.Kucuk[e], T.Kurtukian-Nieto[c], F.Molina[a,h], L.Popescu[i], A.M.Rogers[j], G.Susoy[e], C.Stodel[f], T.Suzuki[g], A.Tamii[g] and J.C.Thomas[f]

[a]*Instituto de Física Corpuscular, CSIC-Universidad de Valencia, E-46071 Valencia, Spain*
[b]*Department of Physics, Osaka University, Toyonaka, Osaka 560-0043, Japan*
[c]*Centre d'Etudes Nucléaires de Bordeaux Gradignan, Université Bordeaux 1, Gradignan, France*
[d]*Department of Physics, University of Surrey, Guilford GU2 7XH, Surrey, UK*
[e]*Department of Physics, Istanbul University, Istanbul, Turkey*
[f]*Grand Accélérateur National d'Ions Lourds, BP 55027, F-14076 Caen, France*
[g]*Research Center for Nuclear Physics, Osaka University, Ibaraki, Osaka 567-0047, Japan*
[h]*Comisión Chilena de Energía Nuclear, Casilla 188-D, Santiago, Chile*
[i]*Vakgroep Subatomaire en Stralingsfysica, Universiteit Gent, B-9000 Gent, Belgium*
[j]*Physics Division, Argonne National Laboratory, Argonne, Illinois 60439, USA*

**Abstract.** The half-lives of the $T_Z = -2$, $^{56}$Zn and $T_Z = -1$, $^{58}$Zn isotopes and other nuclei were measured in a β-decay experiment at GANIL. The energy levels populated by the $^{56}$Zn β decay were determined. The $^{56}$Zn results are compared with the results of the mirror process, the charge exchange reaction $^{56}$Fe($^{3}$He,$t$)$^{56}$Co.

**Keywords:** β decay. Gamow-Teller transitions. Charge-exchange reactions. Isospin symmetry. Proton-rich-nuclei. $^{56}$Zn. $^{58}$Zn. Decay by proton emission.

## B(GT) DISTRIBUTIONS IN β DECAY AND CE REACTIONS

Beta (β) decay is a powerful tool to investigate the structure of exotic nuclei, giving direct access to the absolute Gamow-Teller transition strengths $B_j$(GT):

$$B_j(\text{GT})^\beta = \frac{K}{\lambda^2} \frac{I_\beta^j(E_j)}{f(Q_\beta - E_j, Z) T_{1/2}} = \frac{K}{\lambda^2} \frac{I_\beta^j(E_j)}{f_j T_{1/2}} \qquad (1)$$

where $T_{1/2}$ is the parent half-life, $I^j_\beta(E_j)$ is the β-feeding to the state at energy $E_j$ in the daughter nucleus, $f(Q_\beta - E_j, Z)$ is the Fermi factor and $K/\lambda^2$ is a constant [1].

The finite $Q_\beta$ value lets only access to states at low excitation energy in β decay, while Charge-Exchange (CE) reactions populate states up to high excitation energies. At intermediate beam energies and zero momentum transfer ($\vartheta \sim 0°$), the CE differential cross-sections for GT transitions are closely proportional to $B_j$(GT) [2]:

$$\left[ \frac{d\sigma_{GT}^{CE}}{d\Omega}(0°) \right]_j \cong \hat{\sigma}_{GT}(0°) B_j(\text{GT}) \qquad (2)$$

allowing the determination of relative $B_j$(GT) values. Under the assumption of isospin symmetry, β decay of proton-rich nuclei and CE reactions on the mirror stable target nuclei complement each other. Absolute GT strengths up to high excitation energies are extracted by the combined analysis of the mirror processes [1,3]:

$$\frac{1}{T_{1/2}} = \frac{1}{K}\left[B(\text{F})(1-\delta_C)f_F + \sum_{j=GT}\lambda^2 B_j(\text{GT})f_j\right] \quad (3)$$

where the Fermi strength $B(\text{F}) = |N - Z|$, and $f_F$ and $f_j$ can be calculated if the decay energy is known. The relative strengths proportional to $B_j$(GT) are obtained by the CE ($^3$He, $t$) reaction and the absolute normalization is given by the total half-life $T_{1/2}$ of the β decay. Thus, a precise value of the parent half-life is very important.

Based on these ideas, a series of β-decay experiments and CE reactions starting from mirror nuclei were carried out. The $\beta^+$ decays of the $T_Z = -1$ $^{42}$Ti, $^{46}$Cr, $^{50}$Fe and $^{54}$Ni nuclei were studied at GSI [1,4,5]. The results motivated further investigations on more exotic $fp$-shell nuclei, the $T_Z = -1$, $^{58}$Zn and $T_Z = -2$, $^{56}$Zn isotopes, which have been studied in two separate experiments at GANIL. The $\beta^-$-type mirror CE reactions on the $T_Z = +1$, $^{58}$Ni and $T_Z = +2$, $^{56}$Fe target nuclei were carried out at RCNP Osaka [1,3] using the ($^3$He,$t$) reaction at 140 MeV/u and $\vartheta = 0°$.

## Experimental Set-up and Half-life Analysis

The experiment performed in 2010 at the LISE3 facility of GANIL [6] was focused on $^{56}$Zn. Data was also taken for $^{58}$Zn, a nucleus of astrophysics interest since it constitutes a waiting point in the *rp*-process. A 74.5 MeV/u $^{58}$Ni$^{26+}$ primary beam with average intensity of 3.7 eμA was fragmented on a 200 μm $^{nat}$Ni target. The fragments were selected by the ALPHA-LISE3 separator and implanted into a Double-Sided Silicon Strip Detector (DSSSD), surrounded by four Ge EXOGAM clovers for gamma (γ) detection. The 300 μm thick DSSSD, with 16×16 strips and a pitch of 3 mm, was used both as an implantation detector and to detect the subsequent β decays, using two parallel electronic chains with different gains. The implanted nuclei were identified on an event-by-event basis thanks to the measurement of the energy loss $\Delta E$, residual energy and Time-of-Flight (*TOF*). Any implantation event triggering the first $\Delta E$ silicon detector was acquired. Decay events were triggered by a signal above threshold (typically 50-90 keV) in the DSSSD, with no coincident signal in the first $\Delta E$ detector.

The time interval between an implant event and the corresponding β-decay event is related to the half-life. Therefore, we performed a time correlation over a period of 50 s: each β-decay event is correlated with all the previous implants happened in the same pixel of the DSSSD. Truly correlated events accumulate in the exponential decay curve, while random correlations form a constant background (see, e.g., Fig. 1). The half-life is then obtained by a least squares fit to the resulting decay-time spectrum, assuming an exponential decay curve for the isotope of interest, growth and decay curves for all the daughter nuclei and a constant background.

## Results for $T_Z = -1$ Nuclei Populated in the $^{58}$Zn Setting

The $T_Z = -1$ nuclei, obtained with magnetic settings optimized for $^{58}$Zn, have no or very weak β-delayed proton (p) emission. The β decay of the daughter leads to a stable or long-lived nucleus, thus the half-life fit only includes parent decay and daughter growth and decay. The half-lives measured for the $T_Z = -1$ nuclei, given in Table 1, are in agreement with the results of a previous GANIL experiment [7] and the literature values.

**TABLE 1.** Experimental half-lives of $T_Z = -1$ proton-rich nuclei produced in the $^{58}$Zn setting.

| Isotope | $T_{1/2}$ [ms] | Number of implants |
|---|---|---|
| $^{58}$Zn | 88 ± 5 | 82138 |
| $^{56}$Cu | 82 ± 2 | 554374 |
| $^{54}$Ni | 115 ± 4 | 263497 |
| $^{52}$Co | 124 ± 23 | 19483 |

## Results for the $T_Z = -2$ $^{56}$Zn Isotope

The decay of $T_Z = -2$ nuclei is more complex than the $T_Z = -1$ case, because both β-delayed γ rays and β-delayed p-emission are present. Due to the low proton separation energy in $^{56}$Cu, $S_p = 560$ keV, all the levels populated in the daughter decay via p-emission. We assume that the only possible γ-branching is from the decay of the Isobaric Analog State (IAS), since the p decay of the IAS is isospin forbidden (although it has been observed [8]). Thus, the fit of the decay-time spectrum includes the $^{56}$Zn decay, the growth and decay of both the $^{56}$Cu β-daughter and the $^{55}$Ni β-p-daughter, and the β-detection efficiency ($\varepsilon_\beta \sim 12$ % determined in the $^{58}$Zn setting). From the fit shown in Fig. 1, $T_{1/2} = (29.5 \pm 1.0)$ ms was obtained for $^{56}$Zn.

The $^{56}$Zn charged-particle decay-energy spectrum measured by the DSSSD is shown in Fig. 2a. In this spectrum, the background due to the random correlations was removed according to the method of [8], i.e., two decay-energy spectra were created for different cuts of the decay-time spectrum (0-1 s and 1-2 s) and then these spectra were subtracted. Five proton peaks are observed at energies (in $^{56}$Cu excitation energy) that agree well with those of the mirror $^{56}$Co levels from the $^{56}$Fe($^3$He,$t$)$^{56}$Co, CE spectrum (Fig. 2b, reprinted from [9]). The IAS is clearly identified (as in [8]) and all the dominant transitions are observed in both the β decay and the mirror CE spectra, indicating a good isospin symmetry.

The γ-ray spectrum for the decay of $^{56}$Zn is shown in Fig. 3a, where the background of random correlations was subtracted just as for the proton spectrum. A γ ray at 1834 keV is seen, corresponding to the γ decay of the IAS. The half-life analysis for (β-γ)-implant correlations, selecting the γ line observed at 1834 keV, gives $T_{1/2} = (27 \pm 8)$ ms, in agreement with the value from β-implant correlations (Fig. 1).

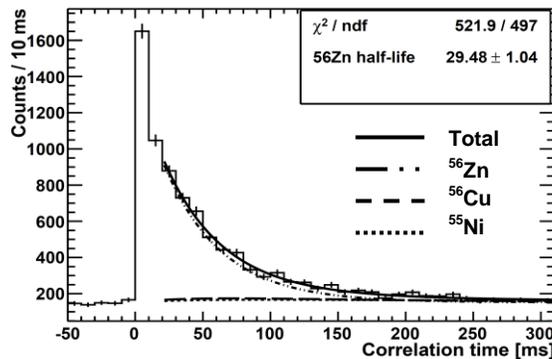

**FIGURE 1.** Decay-time spectrum of $^{56}$Zn. N° of implants = 8861, $T_{1/2} = (29.5\pm1.0)$ ms.

The $^{56}$Zn decay scheme is given in Fig. 3b. It should be noted that both the γ de-excitation and p-emission from a $T = 2$, IAS in a nucleus above the $f_{7/2}$-shell are observed for the first time.

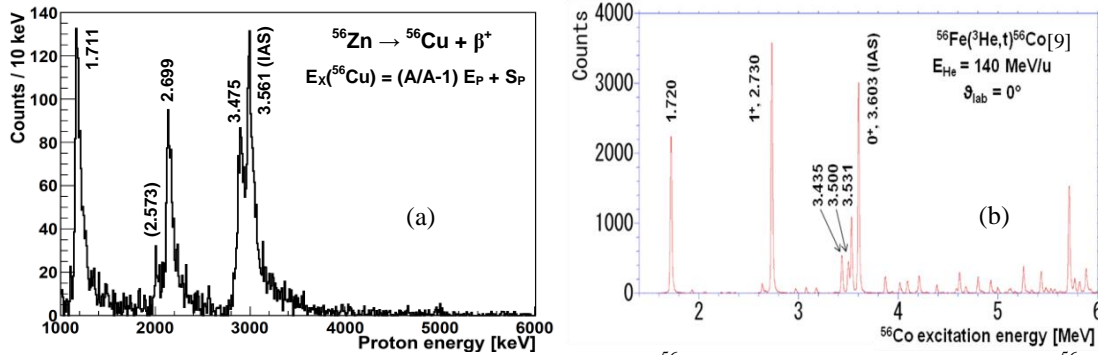

**FIGURE 2.** a) Charged-particle spectrum from the decay of $^{56}$Zn. b) Mirror spectrum from CE on $^{56}$Fe.

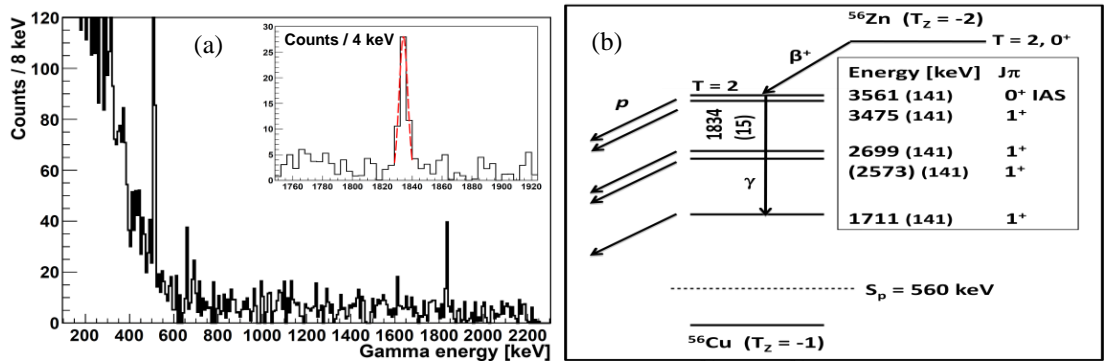

**FIGURE 3.** a) Gamma-ray spectrum for the decay of $^{56}$Zn. b) $^{56}$Zn decay scheme.

# ACKNOWLEDGMENTS

This work was supported under the Spanish MICINN grants FPA2008-06419-C02-01, FPA2011-24553; CPAN Consolider-Ingenio 2010 Programme CSD2007-00042; MEXT, Japan 18540270; and Japan-Spain collaboration program of JSPS and CSIC.